\documentclass{report}
\usepackage[dvips]{graphicx}

\begin{document}

\centerline{\LARGE{Long-term dynamics of Methone, Anthe and Pallene}}
\vspace{0.25cm}
\centerline{\Large{N. Callegari Jr.$^{1}$ \& T. Yokoyama$^{2}$}}
\vspace{0.5cm}\centerline{Departamento de Estat\'{i}stica, Matem\'{a}tica Aplicada e Computa\c{c}\~{a}o} 
\centerline{Universidade Estadual Paulista J\'{u}lio Mesquita Filho (UNESP)}
\centerline{Avenida 24-A no. 1515, CEP 13506-700, Rio Claro/SP/Brazil}
\centerline{email: {\tt $^{1}$calleg@rc.unesp.br, $^{2}$tadashi@rc.unesp.br}}

\vspace{2cm}

\vspace{2cm}

\centerline{\large{ABSTRACT}}

We numerically investigate the long-term dynamics of the Saturn's small satellites Methone (S/2004 S1), Anthe (S/2007 S4) and Pallene (S/2004 S2). In our numerical integrations, these satellites are disturbed by non-spherical shape of Saturn and the six nearest regular satellites. The stability of the small bodies is studied here by analyzing long-term evolution of their orbital elements.

We show that long-term evolution of Pallene is dictated by a quasi secular resonance involving the ascending nodes ($\Omega$) and longitudes of pericentric distances ($\varpi$) of Mimas (subscript 1) and Pallene (subscript 2), which critical argument is $\varpi_2-\varpi_1-\Omega_1+\Omega_2$. Long-term orbital evolution of Methone and Anthe are probably chaotic since: i) their orbits randomly cross the orbit of Mimas in time scales of thousands years); ii) numerical simulations involving both small satellites are strongly affected by small changes in the initial conditions.
\newpage
\section*{1. Introduction}

Short-term dynamics (i.e., time scales of a few years), and determinations of orbital elements of Methone and Pallene, are reported in Porco et al. (2005), Spitale et al. (2006), Porco et al. (2007), Jacobson et al. (2008). Anthe's orbit, the most recent small body detected in Saturnian system, is studied in Cooper et al. (2008).

In spite of some recent investigations (Porco et al. 2005, Callegari \& Yokoyama 2008), long-term numerical integrations (i.e, in time scales of millennia), of the orbits of the three small satellites are not reported yet in the literature. Ephemeris of Methone, Anthe and Pallene are limited to 1,000 years (\emph{Horizons data system}, http://ssd.jpl.nasa.gov). Moreover, inspection of different references listed above show slightly different values for some elements (in particular, the semi-major axis of the small satellites).

In this work, we analyze the results of a great deal of numerical integrations over 60,000-years of the orbits of small satellites similar to Methone, Anthe and Pallene, which initial semi-major axes are varied within the range of different values published in literature.

We have considered the canonical set (Hamiltonian form) of equations of motion within the domain of general N-body problem (e.g. Ferraz-Mello et al. 2005):

\begin{equation}
H=H_0+H_1+H_{\texttt{J}}
\label{1} \\
\end{equation}

\begin{equation}
H_0=\sum_{i=1}^N\left( \frac{|\stackrel{%
\rightarrow }{p_i}|^{2}}{2\beta_i}-\frac{\mu_i\beta_i}{|\stackrel{%
\rightarrow }{r}_i|}\right),
\label{2} \\
\end{equation}

\begin{equation}
H_1=\sum_{0<i<j}\left(-\frac{Gm_im_j}{\Delta_{ij}}+\frac{%
\stackrel{\rightarrow}{p_{i}}\cdot \stackrel{\rightarrow
}{p_{j}}}{M}\right),\hspace{0.2cm} j=1,...,N,
\label{3} \\
\end{equation}

\begin{equation}
H_J=-\sum_{i=1}^N\frac{\mu_i\beta_i}{|\stackrel{%
\rightarrow }{r}_{i}|}\left[-\sum_{l=2}^\infty
J_{l}\left(\frac{R_e}{|\stackrel{%
\rightarrow }{r}_i}\right)^lP_l(\sin\varphi_i)\right],
\label{4} \\
\end{equation}
where $\stackrel{\rightarrow}{r_i}$ and $\stackrel{\rightarrow}{p_i}$ (canonical variables), are position vectors of the satellites relative to the
center of the planet and momentum vectors relative to the center of mass of the system, respectively.
$\mu_i=G(M+m_i)$, $\beta_i=\frac{Mm_i}{M+m_i}$, $\Delta _{ij}=|\stackrel{\rightarrow }{r_i}-\stackrel{\rightarrow }{r_j}|$, where $G$ is the gravitational constant, and $M$, $m_i$ are the planet mass and the individual masses of satellites, respectively. The chosen units are the equatorial radius of the planet ($R_e=60,268$ km), day and $M$.

Equation (\ref{2}) defines the Keplerian motion of each satellite around the planet and Eq.
(\ref{3}) gives the mutual interaction among the satellites. Equation (\ref{4}) represents the perturbation of the non-sphericity
of the planet, $P_{l}(\sin\varphi_{i})$ are the classical Legendre polynomials, $\varphi_{i}$ are the latitudes of the orbits of the satellites referred to the \emph{equator} of the planet. We consider $l=2$, $4$, where in Eq. (\ref{4}) $J_{2}$, $J_{4}$ are the zonal oblateness coefficients.

Equations of the motion of the satellites, which are solved numerically using RA15 code (Everhart 1985), are

\begin{equation}
\frac{d\stackrel{\rightarrow }{r_i}}{dt}=\frac{\partial
H}{\partial \stackrel{\rightarrow }{p_i}},
{\hspace{0.1in} }\frac{d\stackrel{%
\rightarrow }{p_i}}{dt}=-\frac{\partial H}{\partial
\stackrel{\rightarrow }{r_i}}, \hspace{0.25cm}i=1, ..., N,  \label{5}
\end{equation}

All outputs shown in this work are \emph{planetocentric} ones. In the simulations which we show here, we have neglected the mutual perturbations between the small bodies, and numerical simulations include a small satellite and six inner regular satellites (Mimas, Enceladus, Tethys, Dione, Rhea and Titan) (i.e., $N=7$ in all equations above). However, we have tested the effects of the mutual perturbations between the small satellites: the masses of these last bodies are very small and the simulations considering the whole system, clearly confirmed that their mutual effects are negligible.

Initial \emph{osculating} planetocentric elements and parameters of \emph{regular} satellites, and Saturn gravity field, are obtained from Jacobson et al. (2006) and \emph{Horizons data system} (data January 1, 2007). Initial semi-major axis of the small bodies will be indicated in the figures given in next section. In our code, the mass of the small body is equal to the mass of a spherical body with density similar to the Mimas's and $\sim$3 km in diameter (e.g. Porco et al. 2005, Porco et al. 2007).

\begin{figure}[htb] 
       \begin{minipage}[b]{0.48 \linewidth}
           \fbox{\includegraphics[width=\linewidth]{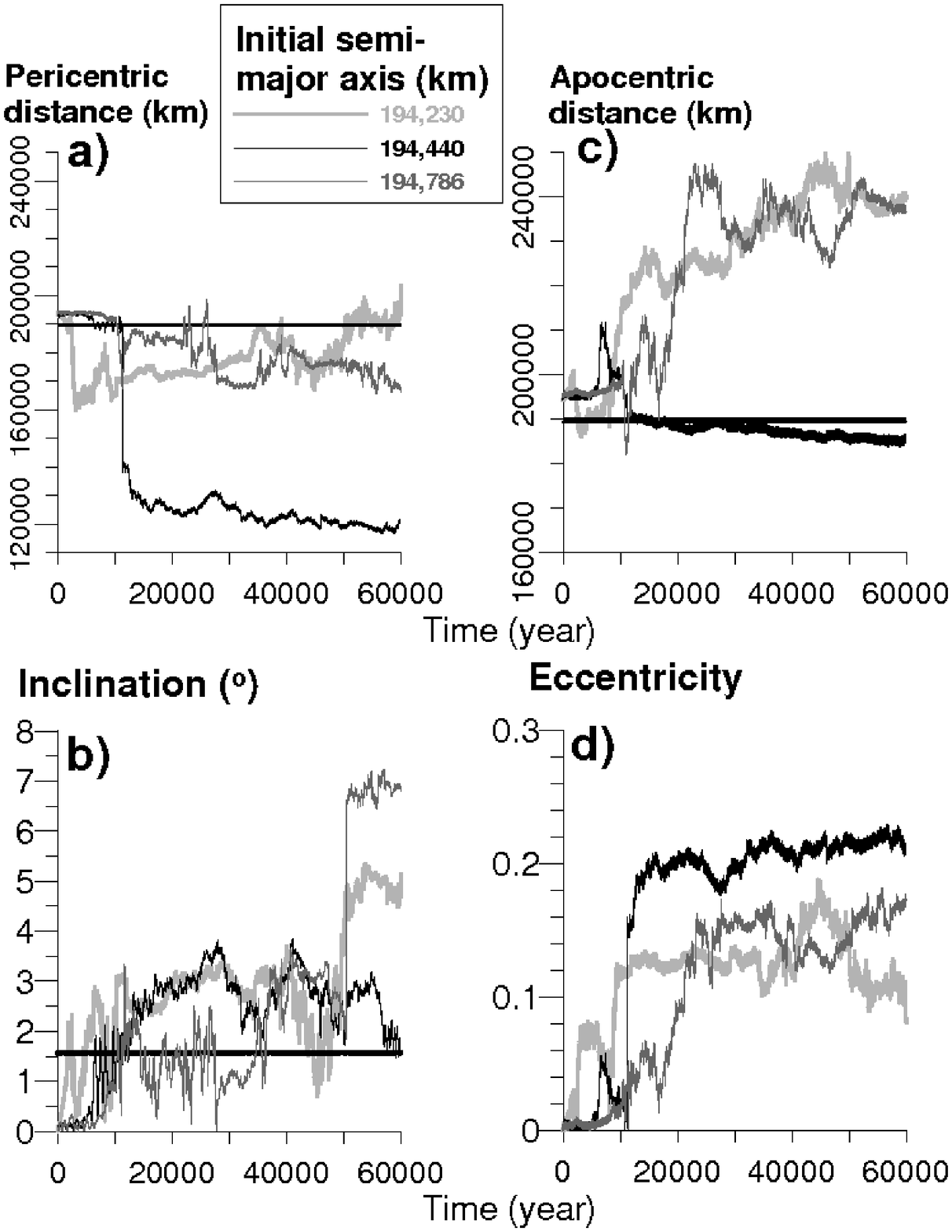}}\\
           \caption{\it Time variation of different quantities (indicated in the plots) of orbits of small satellites similar to Methone. All Methone's initial elements except the initial semi-major axis have been taken from \cite[Jacobson et al. 2008]{Jac2008}. Apocentric distance of Mimas and its inclination are shown by ``horizontal'' curves in a), b), c).}
           \label{fig:XXX}
       \end{minipage}\hfill
       \begin{minipage}[b]{0.48\linewidth}
           \fbox{\includegraphics[width=\linewidth]{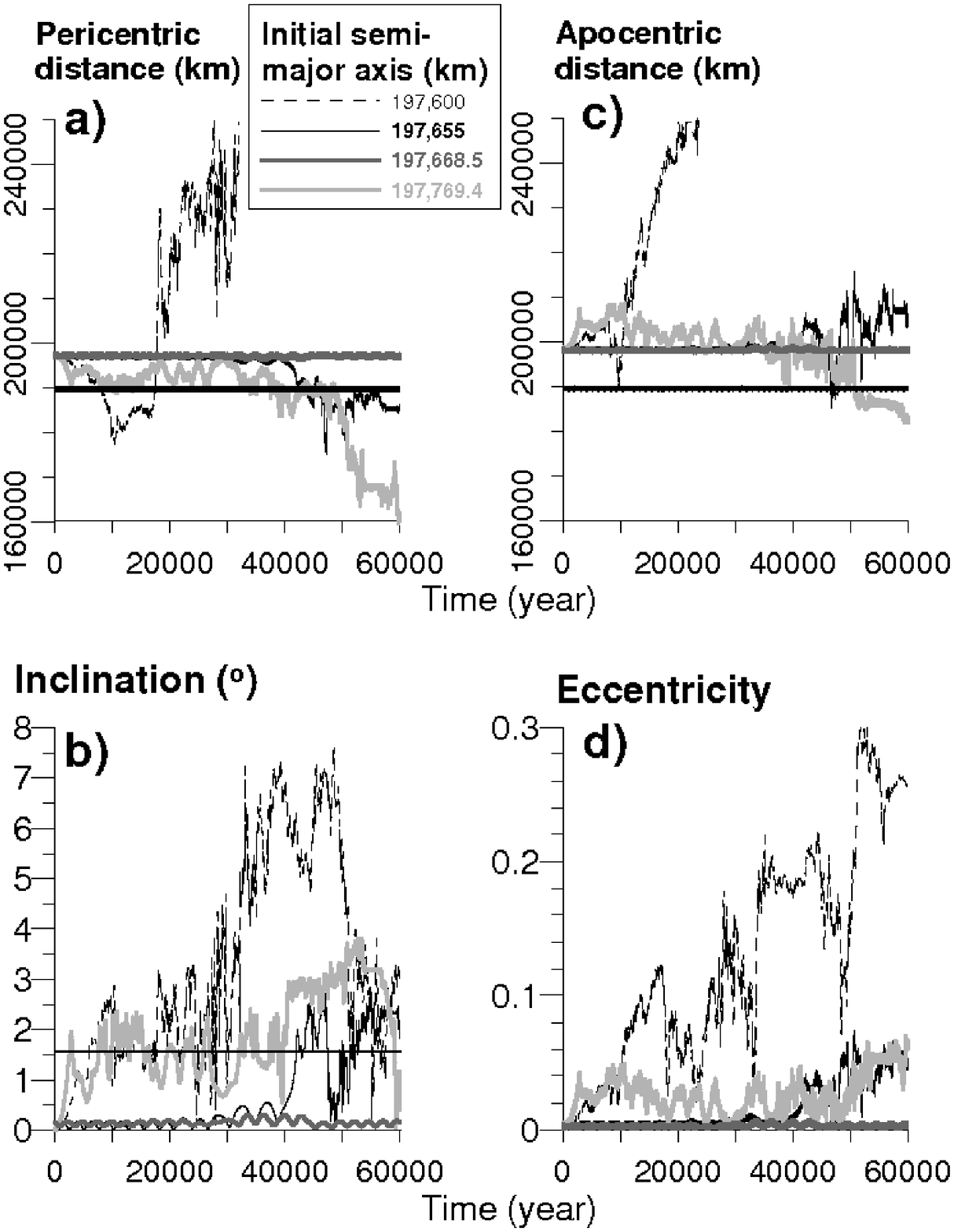}}\\
           \caption{\it Similar to Fig. 1, but for clones of Anthe. Except for the semi-major axis, all Anthe's initial conditions have been obtained from osculating elements given in \cite[Cooper et al. 2008]{Coo2008}. The only exception is the strong-gray curves, where mean values of \cite[Cooper et al. 2008]{Coo2008} are used instead of osculating elements.}
           \label{fig:XXXX}
       \end{minipage}
   \end{figure}

\begin{figure}[htb] 
       \begin{minipage}[b]{0.48 \linewidth}
           \fbox{\includegraphics[width=\linewidth]{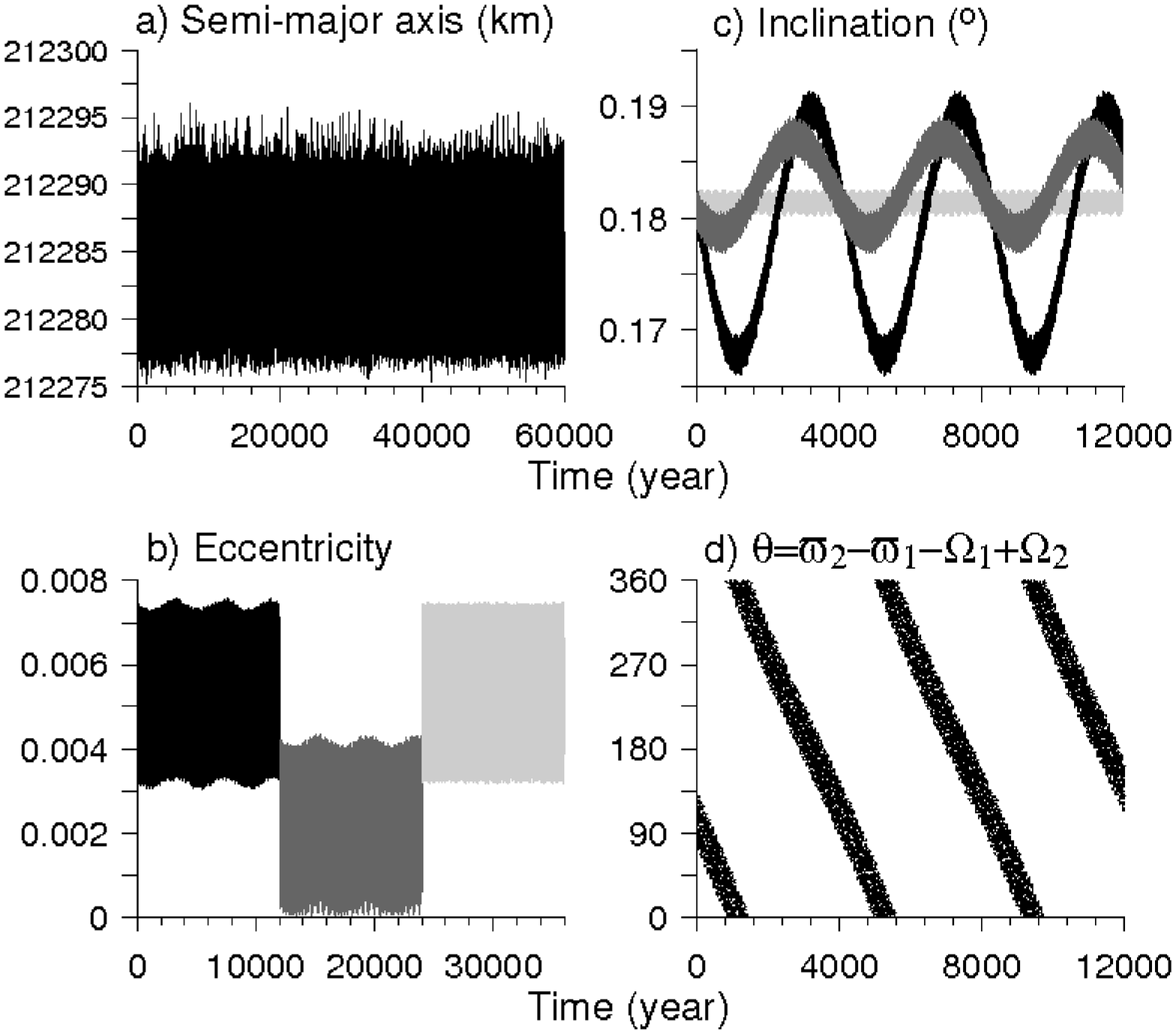}}\\
           \caption{\it Black curves: time variation of different quantities (indicated in the plots) of orbits of small satellites similar to Pallene. $\theta$ is the critical angle associated to quasi-resonance. Except the initial eccentricity in strong-gray curves (where initial eccentricity is null), all initial elements of Pallene have been taken from \cite[Jacobson et al. 2008]{Jac2008}. Light-gray curves: Mimas' mass is set as almost zero in the code. In b), the different portions of the eccentricity variation of the three simulations are shown.}
           \label{fig:XXX}
       \end{minipage}\hfill
       \begin{minipage}[b]{0.48 \linewidth}
           \fbox{\includegraphics[width=\linewidth]{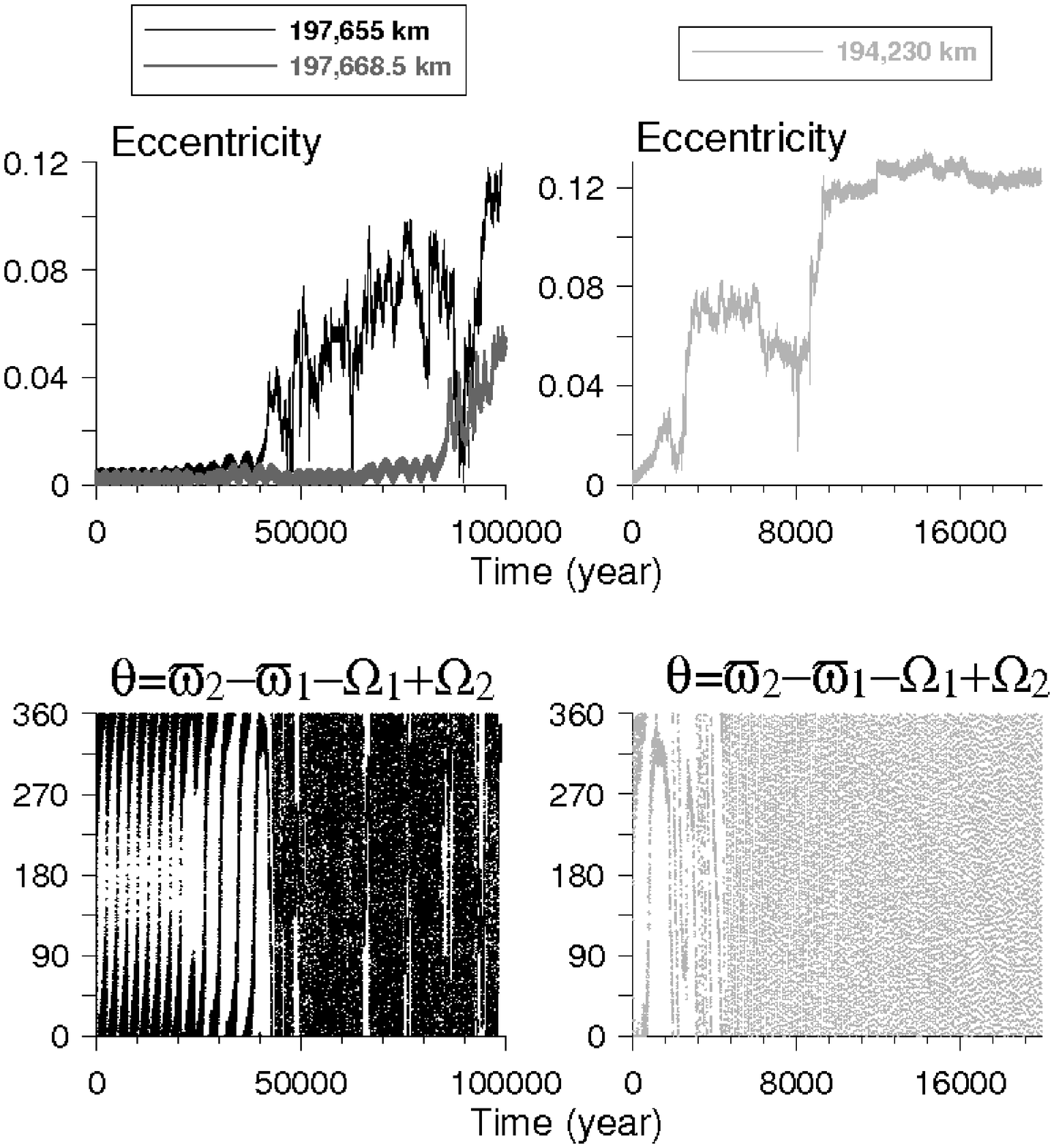}}\\
           \caption{\it Time variation of eccentricity and critical angle of some orbits with same colors given in Figure 1 (right column, Methone) and Figure 2 (left column, Anthe), in different time scales than those shown in Figs. 1, 2. Note that ``stable'' strong-gray curve in Fig. 2(d) suffers large variation on larger time scales.}
           \label{fig:XXXX}
       \end{minipage}
   \end{figure}

\section*{2. Results}
In this section, we discuss the results of our numerical simulations on long-term evolution of orbits of Methone, Anthe and Pallene. Simulations are shown in the time span of 60,000 years, although several of them were continued until 100,000 years. More than ten different simulations for each satellite have been done with our Fortran code utilizing different desktop computers, operational systems (Windows and Linux), and step-size in the integrations.

\subsection*{2.1 Chaotic Methone?}



In Fig. 1(a), the black (almost horizontal) line is the apocentric distance of Mimas during the time integration of one simulation. The remaining  curves are the pericentric distances for different clones of Methone. Fig. 1(b) shows that the orbital inclination of fictitious Methone varies quite randomly and in particular, during the 60,000 years, Mimas and Methone can share the same orbital plane several times. Therefore a collision seems to be a possibility in this scenario, unless a favorable mutual inclination of these satellites is always guaranteed.

The values of initial Methone's semi-major axis $194,230$ km and $194,440$ km (light-gray and black curves in Fig. 1, respectively), are given by Jacobson et al. (2008) and Spitale et al. (2006), respectively. Porco et al. (2007) lists the same value of Spitale et al. (2006) for the semi-major axis of Methone. Note the large divergence between their corresponding trajectories. The largest variation of eccentricity occurs when we adopt the value $194,440$ km (Fig. 1(d)). The high sensitivity of the evolution of the orbits with the initial conditions is probably an indication of the long-term chaotic motion of Methone. In Callegari \& Yokoyama (2009) we are studying in details the dynamics of Methone with the technique of dynamical maps, where we show that Methone is close to the collision curve with Mimas, its main disturber.

Figure 1(c) shows that the apocentric distances attain in general larger values: in all cases, the quantity attains the Pallene's semi-major axis, ($\sim212,280$) km; in some cases the orbit of Enceladus may be crossed ($\sim 238,408$ km).

\subsection*{2.2 Chaotic Anthe?}
In Fig. 2(a), the listed Anthe's semi-major axes are slightly smaller than $197,770$ km given in Porco et al. (2007).

Curves in Fig. 2(a) show that pericentric distances of all but one fictitious Anthe are smaller than the apocentric distance of Mimas (black trajectories in Figs. 2(a,b)). The discussion above on the possibility of collisions between Methone and Mimas is also valid here. In the case of Anthe, if the initial conditions of the simulations are the osculating elements as given in Cooper et al. (2008), this satellite seems to be stable during the adopted time span (strong-gray curves in Fig. 2). However, for longer times as shown in Fig. 4, this scenario is not maintained.

It is remarkable the effect caused when a small change is applied in the initial semi-major axis of Anthe given in Cooper et al. (2008) ($a=197,655$ km). For instance, for $a=197,600$ km, Anthe's eccentricity attains large values (Fig. 2(d)), so that this satellite can penetrate  Dione's orbit, crossing also the orbits of Tethys and Enceladus (see dashed curves are interrupted in order to keep the scale of y-axes).

The previous discussion on probable long-term (chaotic) instability  of Methone is also a possibility in the case of Anthe, though its orbit is slightly far from collision curve with Mimas (Callegari \& Yokoyama 2009).


\subsection*{2.3 Quasi secular resonance between Mimas and Pallene}

Black curves in Fig. 3 show two typical results taken from several long-term numerical simulations involving small satellites similar to Pallene. Contrary to the cases discussed above, now the semi-major axes suffer small variations (on the order of 20 km), while eccentricity and inclination also vary between small extremes with a quasi-periodic modulation (not present in semi-major axis evolution).

The oscillations seen in Figs. 3(b,c) are due to proximity of Pallene's orbit of a type of quasi secular resonance associated to the critical angle $\theta=\varpi_2-\varpi_1-\Omega_1+\Omega_2$, which circulates in retrograde sense with period $\sim 4400$ years (Fig. 3(d)).

Porco et al. (2005) pointed out that the non-null value of the current Pallene's eccentricity ($\sim0.004$, similar to the Enceladus'), could be explained by some secular resonance. Here we identify a possible candidate, but it is a \emph{quasi-resonance}, and its effect is not strong enough to increase the eccentricity of the small satellite.

In fact, light-gray curve in Fig. 3(b) shows eccentricity of a small body in a simulation where Mimas' mass has been taken almost zero: though the quasi-periodic oscillations disappears, the interval of variation in eccentricity is almost the same as that seen in the real case, black line. (Annulling Mimas' effects, the variation of inclination however is very small; see Fig. 3(c)).

The variation in eccentricity is a natural outcome due to joint $J_2$  and Mimas' effects. ($J_4$ effects can be negligible; $J_2=0.0162906$ and $J_4=-0.000936$.) Though not shown in Fig. 3, we have tested the individual effects of Enceladus and Titan (annulling their effects on the program): they are not responsible for the variation in eccentricity. Plot in strong-gray in Figs. 3(b,c) show that an initial null eccentricity evolves to a maximum near the current value. So, if initial Pallene's eccentricity were null, currently Pallene is near this maximum.



Some curves presented in Fig. 1 are shown again in Fig. 4. It can be see observed that the jump of the eccentricity seems to occur when the critical angle $\theta$ alternates between circulation and libration. Alternatively we can say that the reason of the transition of $\theta$ is caused by the significative jump suffered by the eccentricity and inclinations which was caused by the close approach Mimas-small satellite. Since this close approach seems to cause more drastic effects that the $\theta$-quasi resonance (which is of the order of $sin(i_1)sin(i_2)e_1e_2$), the alternation circulation-libration is only a consequence and not the reason of the increase of the small satellite's eccentricities and inclinations.

\section*{4. Conclusions}

Complementing previous simulations (Callegari \& Yokoyama 2008), we show here that the orbits of Methone and Anthe are probably chaotic and they can cross Mimas' orbit several times in a few thousand years. The possibility of collision of the satellites with Mimas must be taken into account in the evolutionary studies of the bodies (see Porco et al. (2007) and references therein). The orbit of Pallene is long-term stable in the time scale of the numerical simulations studied here. A quasi secular resonance involving Pallene and Mimas orbits was identified, but its effects are not important to long-term evolution of the small satellite.

Recently (Porco 2009), a new small body (S/2008 S1) was found close to the G-ring. Preliminary 100,000-years numerical simulations with our code show that the orbit of S/2008 S1 is long-term stable, suffering only small variations in the elements ($\Delta$$a=10$km, $\Delta$$e=0.006$, $\Delta$$i=0.004^{\circ}$). It is worth noting that due the to the proximity to the main ring system and its very small size (radius of about 250 meters), long-term dynamics of S/2008 S1 is more complex than their neighbors. 

Acknowledgements: Fapesp (06/58000-2, 06/61379-3, 08/52927-2) and CNPQ.

\end{document}